\documentclass[journal, 10pt]{IEEEtran}

\usepackage{multirow}
\usepackage[font=scriptsize,caption=false,labelsep=space]{subfig}
\usepackage{mathrsfs}
\usepackage{graphicx,float,wrapfig,epstopdf,amsmath}
\usepackage[]{algorithmicx}
\usepackage{algpseudocode,algorithm}
\usepackage{balance}
\usepackage{mathtools,lipsum}
\usepackage{amsmath}
\usepackage{amssymb}
\usepackage{amsthm}
\usepackage{graphicx}
\usepackage{epstopdf}
\usepackage{times}
\usepackage{textcomp,cite}

\usepackage{url}
\usepackage{hyperref}

\usepackage{multirow}
\usepackage{threeparttable}
\graphicspath{{./figures/}}

\usepackage[table]{xcolor}
\definecolor{intnull}{RGB}{213,229,255}
\definecolor{inteins}{RGB}{128,179,255}
\definecolor{color1}{RGB}{199,209,232}
\definecolor{color2}{RGB}{230,231,233}
\definecolor{bg}{RGB}{233,233,227}

\begin{document}

	\title{ Beam-Squint Compensation in ISAC: \\ Causes and Mitigation }
	\title{The Curse of Beam-Squint in ISAC: \\ Causes, Implications, and Mitigation Strategies}
	
	\author{\IEEEauthorblockN{Ahmet M. Elbir, \textit{Senior Member, IEEE,} Kumar Vijay Mishra, \textit{Senior Member, IEEE,}  \\ Abdulkadir Celik, \textit{Senior Member, IEEE,} and Ahmed M. Eltawil, \textit{Senior Member, IEEE}}
		
		\thanks{A. M. Elbir is with 
			King Abdullah University of Science and Technology, Saudi Arabia and Duzce University, Turkey (e-mail: ahmetmelbir@ieee.org).} 
		\thanks{K. V. Mishra is with the United States Army Research Laboratory, Adelphi, MD 20783 USA (e-mail: kumarvijay-mishra@uiowa.edu). } 
		\thanks{A. {C}elik and A. M. Eltawil are with King Abdullah University of Science and Technology, Saudi Arabia (e-mail:  abdulkadir.celik@kaust.edu.sa, ahmed.eltawil@kaust.edu.sa).  }	
	}
	\maketitle
	
	\begin{abstract}
		Integrated sensing and communications (ISAC) has emerged as a means to efficiently utilize spectrum and thereby save cost and power. At the higher end of the spectrum, ISAC systems operate at wideband using  large  antenna arrays to meet the stringent demands for high-resolution sensing and enhanced communications capacity. However, the wideband implementation entails beam-squint, that is, deviations in the generated beam directions because of the narrowband assumption in the analog components. This causes significant degradation in the communications capacity, target detection, and parameter estimation. This article presents the design challenges caused by beam-squint and its mitigation in ISAC systems. In this context, we also discuss several ISAC design perspectives including far-/near-field beamforming, channel/direction estimation, sparse array design, and index modulation. There are also several research opportunities in waveform design, beam training, and array processing to adequately address beam-squint in ISAC. 

	\end{abstract}

	\section{{Introduction}}
	\label{sec:Introduciton}
	\IEEEPARstart{F}{or} several decades, radar and communications systems have been operating exclusively in different bands to minimize the interference to each other. 	While radar systems occupy a large portion of the frequency spectrum, communications systems have moved up from lower frequencies to millimeter-wave (mmWave).	The high frequency operations in both mmWave ($30-300$ GHz) and terahertz (THz)  ($0.1-10$ THz) bands attracted several applications in both radar and communications because they offer ultra-wide bandwidth and extremely high angular/range resolution~\cite{radarCommSurveyfanLiuJSTSP_Zhang2021Sep}. Possible radar applications include automotive radar ($24$-$80$ GHz), indoor localization ($260$ GHz), and cloud observation ($95$ GHz). Meanwhile,  the fifth generation (5G) communications networks employ $28$ GHz as per  third generation partnership project (3GPP) 38.104 standard, and the US Federal	Communications Commission (FCC) has already adopted new	rules for the experimental licenses at $95$ GHz and $3$ THz~\cite{elbir_Beamforming_SPM_Elbir2023Jun,radarCommSurveyfanLiuJSTSP_Zhang2021Sep}. As the operating frequencies converge, \textit{integrated sensing and communications} (ISAC) has emerged to save hardware cost and improve the resource management with a single integrated hardware platform with a common signaling mechanism.
	
	The ISAC paradigm has catalyzed new research opportunities in applications such as autonomous vehicles with high-precision sensing, indoor localization, industrial communications/monitoring, and extended/virtual reality. These applications demand seamless communications performance with other users/vehicles/infrastructures as well as radar sensing with high angular/range resolution. To meet these requirements, the ISAC systems harness large number of antenna elements while occupying ultra-wide bandwidth to achieve high spectral efficiency (SE) as well as pencil beamforming to complement both sensing and communications (S\&C) functionalities. In this context, the use of hybrid analog/digital beamformers is a common choice to generate spatially diverse beams toward users/targets and maintain cost/hardware efficiency~\cite{elbir_Beamforming_SPM_Elbir2023Jun}. 
	
\begin{figure*}[t]
		\centering
		{\includegraphics[draft=false,width=.8\textwidth]{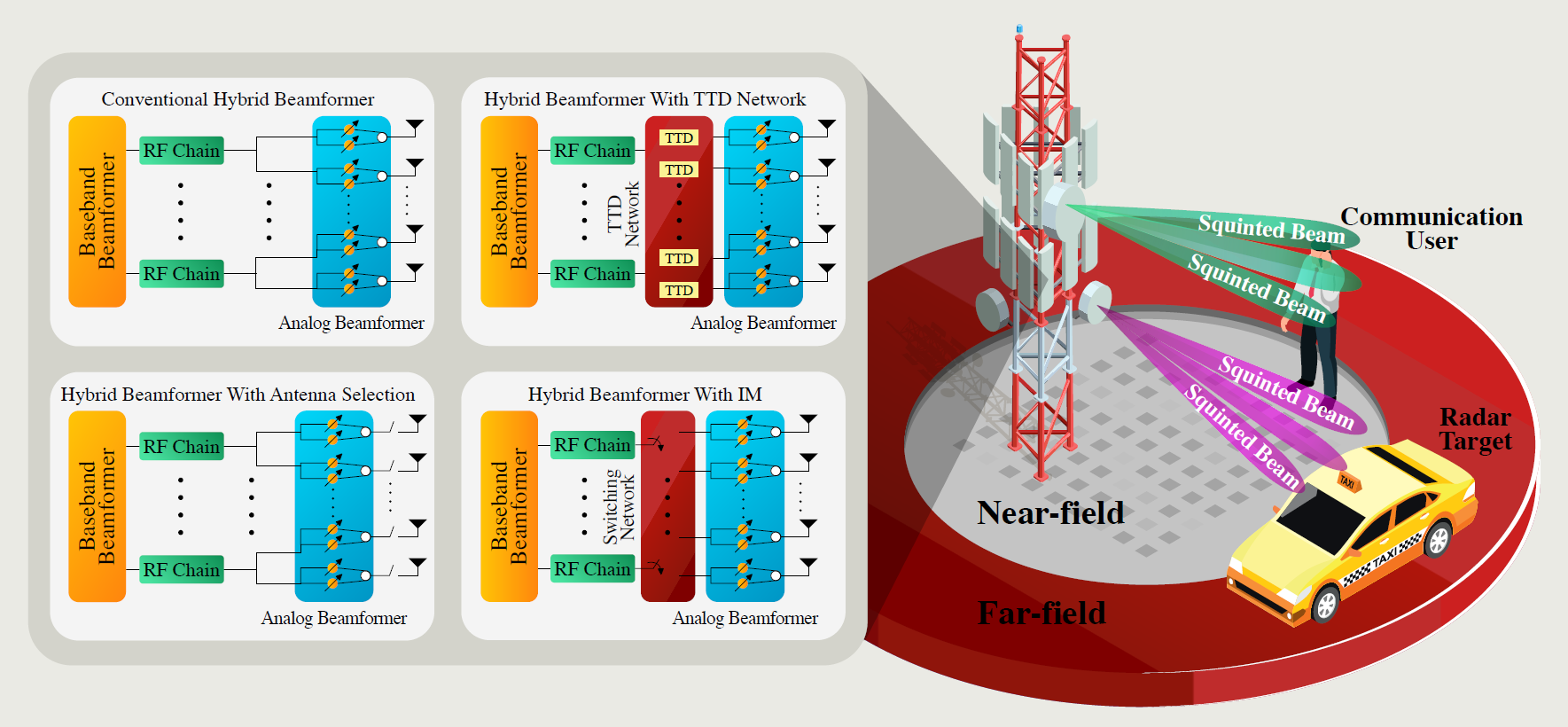} }
		\caption{ISAC beam generation in the presence of beam-squint with various transmitter architectures based on conventional hybrid analog/digital beamformers incorporated with TTD networks, antenna selection, and spatial path IM. 
		} 
		\label{fig_diag}
	\end{figure*}

	{The hybrid beamforming architectures perform subcarrier-dependent baseband processing while the analog beamformers are subcarrier-independent. That is, the same analog beamformer weights, which are usually set according to the carrier frequency, are adopted for all subcarriers. Thus, when the frequency separation among the subcarriers is large, the beams generated by different subcarriers (especially at the low/high-end frequencies) squint, and they illuminate distinct directions, as illustrated in Fig.~\ref{fig_diag}. This phenomenon is called \textit{beam-squint}.
		In particular, the deviation from the carrier frequency towards the high-end and low-end subcarriers causes forward and backward beam-squints, respectively~\cite{ttd_mmWave_Lin2022}. The beam-squint leads to severe degradation in the communications capacity, as well as deterioration of the target detection and estimation accuracy. In addition, the impact of beam-squint becomes more severe in large arrays as the beam-width becomes much narrower.} For instance, consider two users/targets located in the far- and near-field at the broadside direction (that is, perpendicular to the antenna array plain). In the far field, the angular deviation due to beam-squint is roughly $6^\circ$ ($0.4^\circ$) for $300$ GHz with $30$ GHz ($60$ GHz with $1$ GHz) bandwidth, respectively~\cite{ttd_mmWave_Lin2022}. However, near-field beam-squint has approximately $(6^\circ, 10 \text{m} )$ angular and range deviation for a target/user located at $20$ m distance for $300$ GHz with $30$ GHz bandwidth~\cite{nearField_ISAC_Wang2023May,elbir_Beamforming_SPM_Elbir2023Jun}. 


	In S\&C applications, beam-squint compromises the reliability of communications links and the deterioration of target detection/tracking accuracy. Moreover, it leads to the reduction of instantaneous bandwidth and beam-forming gain, as well as signal-to-noise ratio (SNR)~\cite{beamSquintAware_Satellite_You2022Aug,delayPhasePrecoding_THz_Dai2022Mar}. Accurate direction-of-arrival (DoA) estimation and localization of the targets are required for high-precision sensing applications. Furthermore, spectral efficiency (SE) is maximized by accurately handling beam-steering in massive multiple-input multiple-output (MIMO) communications, wherein the spatial diversity is utilized. As a result, the ISAC system design should take into account the impact of beam-squint to achieve high-precision sensing and enhanced communications capacity~\cite{trueTimeDelayBeamSquint}. 
	
	Due to the frequency-dependent nature of beam-squint, its compensation should be carried out at all subcarriers for the operating bandwidth. The state-of-the-art techniques include hardware-based (e.g., true-time delayer (TTD))~\cite{delayPhasePrecoding_THz_Dai2022Mar,beamSquintAware_Satellite_You2022Aug}  or signal processing-based~\cite{elbir_thz_SPIM_ISAC_Elbir2023Mar,elbir_MC_beamSquint_DOA_Elbir2023Sep} approaches. While there is no unique and ultimate solution to perfectly eliminate beam-squint, each solution has its own benefits and drawbacks.

	In this article, we provide an overview of beam-squint compensation in ISAC design. Although the impact of beam-squint is extensively investigated in mmWave~\cite{trueTimeDelayBeamSquint,beamSquint_ISAC_Gao2023Mar} and THz communications~\cite{delayPhasePrecoding_THz_Dai2022Mar,elbir_thz_SPIM_ISAC_Elbir2023Mar,elbir_MC_beamSquint_DOA_Elbir2023Sep}, its implications in ISAC remain relatively unexamined. In the next section, we present state-of-the-art techniques for beam-squint compensation based on TTDs as well as phase correction approaches. Then, we discuss the impact of beam-squint in several ISAC applications varying from beamforming, channel and DoA estimation, antenna selection, and index modulation (IM). Enabling ISAC faces several design challenges, such as waveform design and beam training, TTD design, and array imperfections. These challenges are also discussed in detail along with related future research directions. We believe that this work paves the way for the development of new techniques and algorithms that can facilitate the implementation of wideband ISAC systems. 
	
	\section{{Beam-Squint Compensation Techniques}}
	A common solution to compensate beam-squint is to realize subcarrier-dependent (SD) beams by means of employing additional time-domain components and phase correction techniques~\cite{beamSquintAware_Satellite_You2022Aug}. Time-domain-based techniques include TTD components, which are well-known in radar signal processing applications, to electronically steer the beams without physically moving the antenna array~\cite{ttd_Rotman2016Jan,delayPhasePrecoding_THz_Dai2022Mar}. Another approach is to deal with beam-squint in the digital domain via modifying the digital beamformer weights to account for phase corruptions in the analog beamformer due to beam-squint for each subcarrier without employing additional hardware components~\cite{elbir_thz_SPIM_ISAC_Elbir2023Mar,elbir_MC_beamSquint_DOA_Elbir2023Sep}. Similar to the array designs based on analog, digital or hybrid configurations, different array architectures based on Butler or Blass matrix networks also require calibration for beam-squint in wideband implementation~\cite{elbir_Beamforming_SPM_Elbir2023Jun}.
	
	Beam-squint compensation techniques mainly depend on the underlying antenna feed network architecture; series, parallel, or hybrid. 
	Due to their low cost and complexity, series-fed arrays are mainly used for compact applications such as mmWave radars at $79$ GHz or mmWave communications at $28$ GHz~\cite{ttd_Rotman2016Jan}. Note that most of the commercially available antenna arrays for mmWave communications employ subarrayed structures with hybrid-fed solutions such as array-of-subarrays (AoSAs) and group-of-subarrays (GoSAs) architectures~\cite{elbir_Beamforming_SPM_Elbir2023Jun}. For instance, a $P \times Q$ array typically contains $P$ parallel-fed subarray ports, each of which comprises $Q$ series-fed antennas~\cite{ttd_mmWave_Lin2022}. Beam-squint in series-fed arrays can be overcome by employing reactive circuit elements, e.g., negative capacitor and inductors, to introduce a constant negative group delay, which in turn compensates beam-squint~\cite{ttd_Rotman2016Jan,elbir_MC_beamSquint_DOA_Elbir2023Sep}. Nevertheless, these approaches are mostly favorable to coping with beam-squint at the broadside~\cite{ttd_Rotman2016Jan}. In contrast, parallel-fed arrays, e.g., large arrays, require complicated feed networks such as complex radio-frequency (RF) chains for advanced analog/digital signal processing. Herein, we discuss the state-of-the-art techniques to compensate for beam-squint in ISAC, summarized in Table~\ref{tableSummary}. 
	
	
	\begin{table}
		\caption{State-of-the-Art in Beam-Squint Compensation
		}
		\footnotesize
		\label{tableSummary}
		\centering
		\begin{tabular}{p{0.12\textwidth}p{0.14\textwidth}p{0.16\textwidth}}
			\hline 
			\hline
			\centering\arraybackslash \bf Method  
			&\centering\arraybackslash \bf Advantages
			& \centering\arraybackslash \bf Drawbacks		 \\
			\hline
			\arraybackslash SD Phase Shifter Networks~\cite{beamSquintAware_Satellite_You2022Aug}
			&\arraybackslash   Full compensation of beam-squint in all subcarriers
			& \arraybackslash 	High power consumption and hardware complexity	 \\
			\hline 
			\arraybackslash \centering   TTD Networks~\cite{delayPhasePrecoding_THz_Dai2022Mar,beamSquint_ISAC_Gao2023Mar}
			&\arraybackslash    Easy implementation and  satisfactory performance
			& \arraybackslash Performance  depending on the number of TTD units, time delay range, and resolution		 \\
			\hline
			\arraybackslash Beam Broadening Techniques~\cite{trueTimeDelayBeamSquint}
			&\arraybackslash  No additional hardware components are required
			& \arraybackslash 	Reduced array gain due to subarrayed structure	 \\
			\hline 
			\arraybackslash Phase Correction Techniques~\cite{elbir_MC_beamSquint_DOA_Elbir2023Sep,elbir_thz_SPIM_ISAC_Elbir2023Mar}
			&\arraybackslash  No additional hardware components are required
			& \arraybackslash 	Limited performance depending on the number of RF chains	 \\
			\hline 
			\hline
		\end{tabular}
	\end{table}

	\subsection{{Beam Broadening}}
	A simple approach to reduce the impact of beam-squint is to divide the whole array into multiple subarrays to generate broader beams~\cite{trueTimeDelayBeamSquint}. The generated beams at each subcarrier are now so broad such that they all cover the beam direction corresponding to the carrier frequency, thereby all the beams across the subcarriers seem to be aligned. This approach is also useful since it does not require any structural changes in the transceiver design. However, this reduces the array gain significantly. Furthermore, beam broadening can only be applied in low frequencies, wherein the bandwidth is relatively narrow. In high frequencies, e.g., THz, over ultra-wide bandwidth, the beam directions are so apart from each other that beam broadening cannot work properly.

	\subsection{{TTD Networks}}
	TTD units are composed of switched delay lines with high-precision quantized delays. When placed in phased-arrays, they introduce specific time delays on the input signal. Compared to phase shifters, TTD units can deliver multiple wavelengths of phase shift by virtue of their time-domain operation. Specifically, a TTD unit and a phase shifter consume approximately $100$ mW and $40$ mW, respectively, for THz-band implementation~\cite{ttd_mmWave_Lin2022,elbir_Beamforming_SPM_Elbir2023Jun}. In hybrid beamforming architectures (see Fig.~\ref{fig_diag}), a TTD network is typically positioned ahead of a conventional analog beamformer, which is realized with subcarrier-independent (SI) phase shifters~\cite{delayPhasePrecoding_THz_Dai2022Mar}. 
	
	\subsection{{SD Phase Shifter Networks}}
	The SD beams can also be generated by employing phase shifter networks for each subcarrier such that all the generated beams across the subcarriers look in the same direction~\cite{beamSquintAware_Satellite_You2022Aug}. However, this approach brings the hardware cost of deploying large number of phase shifters. In particular, it requires $MNK$ phase shifters, instead of using only $NK$ phase shifters in conventional hybrid beamforming architectures,  where $M$, $N$, and $K$ denote the number of subcarriers, antennas, and RF chains, respectively.
	
	\subsection{{Phase Correction Techniques}}
	In wideband hybrid arrays, the digital beamformers are SD whereas the analog part is SI. Therefore, it is possible to handle beam-squint compensation in the digital domain instead of employing additional phase shifters or TTDs. This can be done by updating the digital beamformer weights via a linear transformation from the SI phase weights to the SD ones such that the the combined hybrid beamformer achieves the SD beamformer weights~\cite{elbir_thz_SPIM_ISAC_Elbir2023Mar}. While this approach is advantageous since it does not require any additional hardware components, its performance directly depends on the number of RF chains to provide accurate transformation between the inaccurate and corrected phase shifts. Therefore, beam-squint compensation is limited given that fewer RF chains are usually employed in ISAC systems~\cite{elbir_Beamforming_SPM_Elbir2023Jun,radarCommSurveyfanLiuJSTSP_Zhang2021Sep}. 
	
	
	%

	\section{{Beam-Squint in ISAC}}
	{The compensation of beam-squint in ISAC faces several application-specific challenges, such as beamforming, channel/DoA estimation and antenna selection. Herein, we present some physical layer applications for ISAC in the presence of beam-squint, along with an extensive discussion on the existing signal processing techniques.   }
	
	\subsection{{Beamforming}}
	The detrimental impacts of beam-squint on beamforming gain and its potential remedies highly depend on the underlying beamforming technique and transceiver architecture as explained in the sequel. 
	
	\subsubsection{{Digital Beamforming}}
	When the array data is processed via fully digital beamformers, beam-squint is entirely eliminated owing to the SD implementation~\cite{elbir_Beamforming_SPM_Elbir2023Jun,beamSquintAware_Satellite_You2022Aug}. However, this requires dedicated RF chains for each antenna element, thereby increasing the cost and complexity. While fully digital architectures are typically not the preferred choice for commercial communications systems, their usage in military applications with less stringent budgetary constraints is possible~\cite{elbir_MC_beamSquint_DOA_Elbir2023Sep}.

	
	\begin{figure}[t]
		\centering
		{\hspace{5pt}\includegraphics[draft=false,width=0.8
			\columnwidth]{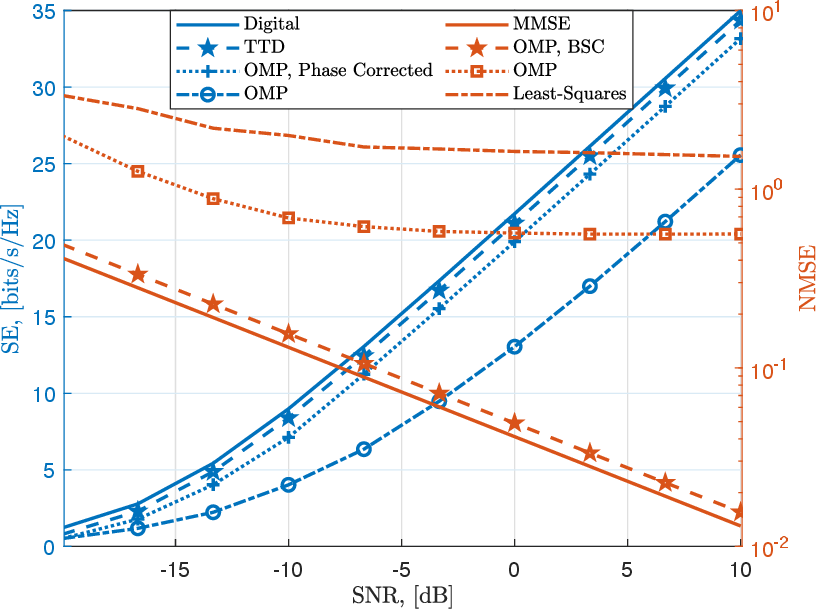} } 
		\caption{Comparison of beam-squint compensation techniques for hybrid beamforming and channel estimation in terms of (left) SE and (right) NMSE when $N=128$, $K=8$, $M=32$  for $300$ GHz carrier frequency and $30$ GHz bandwidth.
		}
		\label{fig_HB_CE}
	\end{figure}
	\subsubsection{{Analog Beamforming}}
	In wideband systems, analog beamforming can be implemented in SD and SI setups. While the SI implementation has a simple hardware architecture with lower beamforming gain because of limited degrees-of-freedom (DoF), the SD analog beamformers can more accurately compensate for beam-squint at the cost of a large number of phase shifter networks~\cite{elbir_thz_SPIM_ISAC_Elbir2023Mar}. Compared to SD phase shifter networks, SD analog beamformer also employs the same number of phase shifters ($MNK$) in the absence of digital part, instead of which simple amplitude controllers may be used~\cite{elbir_thz_SPIM_ISAC_Elbir2023Mar}. As a result, analog architecture exhibits poorer performance compared to digital hybrid beamformers.

	\subsubsection{{Hybrid Beamforming}}
	Hybrid beamforming merges the strengths of both analog and digital beamforming to achieve high performance while mitigating hardware complexity and power consumption. Hence, hybrid architecture exploits much fewer digital components (e.g., baseband processor, analog-to-digital converters, RF chains, etc.) than analog ones (e.g., phase shifters, antenna arrays, etc.)~\cite{elbir_Beamforming_SPM_Elbir2023Jun}.
	
	The beam-squint occurring due to SI analog parts can be compensated by TTD-based delay-phase precoding (DPP) approach~\cite{delayPhasePrecoding_THz_Dai2022Mar} and orthogonal matching pursuit (OMP) with and without phase correction technique~\cite{elbir_Beamforming_SPM_Elbir2023Jun,elbir_thz_SPIM_ISAC_Elbir2023Mar} as depicted in the left axis of Fig.~\ref{fig_HB_CE}, wherein we observe that phase correction technique provides a significant SE performance improvement compared to traditional OMP whereas DPP-TTD with $10$ TTD elements achieves higher SE, and closely follows the digital beamforming performance. Nevertheless, phase correction techniques have the advantage of not using additional hardware at the cost of slight SE loss.


	\subsection{{Channel Estimation}}
	
	
	Consider the receive scenario with conventional hybrid beamformers illustrated in Fig.~\ref{fig_diag}, wherein accurate channel is reconstructed from the received pilot data by employing a virtually generated codebook of SD array responses. These SD array steering vectors are jointly employed to match the received pilot data to account for block-sparsity over subcarriers. Then, the wireless channel can be recovered via sparsity-driven techniques, e.g., OMP, wherein the steering vectors of distinct channel directions for multiple subcarriers are regarded as common blocks~\cite{trueTimeDelayBeamSquint}. While the receive scenario may be handled in the digital domain, the beam-squint compensation techniques should still be employed for the transmit scenario for accurate beamforming. Fig.~\ref{fig_HB_CE} shows the channel estimation accuracy in the right axis in terms of normalized mean-squared error (NMSE) of the estimated channel matrix for the OMP method with and without beam-split compensation (BSC)~\cite{trueTimeDelayBeamSquint,elbir_thz_SPIM_ISAC_Elbir2023Mar}. Large estimation NMSE is observed if beam-squint is not handled accordingly, whereas OMP with BSC attains the minimum MSE performance closely.

	
	\begin{figure}[t]
		\centering{\includegraphics[draft=false,width=.8\columnwidth]{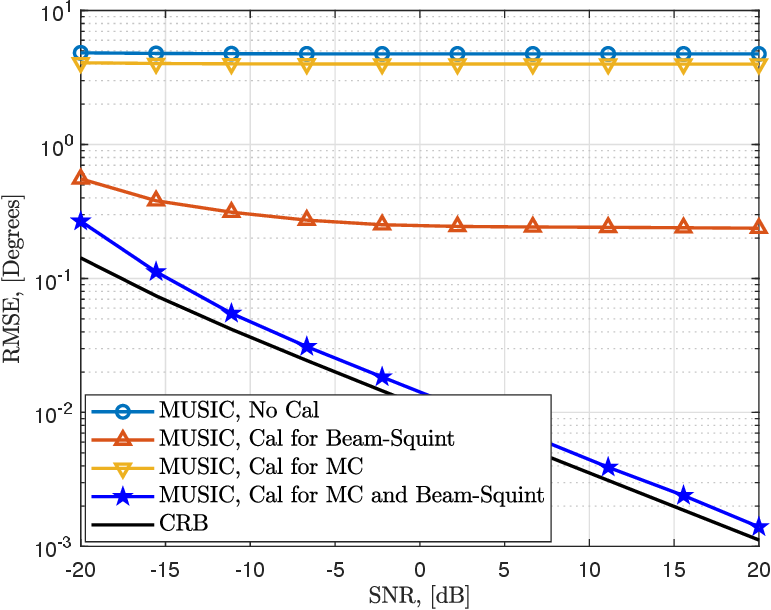}} 
		\caption{DoA estimation RMSE vs. SNR for beam-squint and MC calibration when $N=128$ with $300$ GHz carrier frequency over $30$ GHz bandwidth.
		}
		\label{fig_DOA_SNR}
	\end{figure}

	\subsection{{DoA Estimation}}
	{In radar signal processing, DoA estimation is a well-known problem, wherein usually digital arrays with relatively fewer antennas (up to $16$) are employed compared to MIMO communications~\cite{elbir_Beamforming_SPM_Elbir2023Jun}. With the advances of signal acquisition technologies in MIMO systems, high-precision (e.g., milli-degrees) DoA estimation performance is of great importance for high frequency implementation in mmWave and THz S\&C applications, e.g.,  automotive radar, real-time tracking and user localization~\cite{milliDegree_doa_THz_Chen2021Aug}. In particular, the squinted beams may cause misidentification of the radar targets, thereby degrading the detection and DoA estimation performance in sensing applications. In order examine the impact of beam-squint in sensing, Fig.~\ref{fig_DOA_SNR} presents the DoA estimation root mean-squared-error (RMSE) versus SNR in the presence of both beam-squint as well as mutual coupling (MC) among the antennas. Due to beam-squint, the signal subspace of the array covariance is corrupted. To accurately estimate the DoA angles, a noise subspace correction technique is proposed in~\cite{elbir_MC_beamSquint_DOA_Elbir2023Sep} for the subspace-based approaches, e.g., multiple signal classification (MUSIC) algorithm to account for the deviations due to beam-squint. Beside direction estimation, the range estimation is also effected by beam-squint as it causes the dispersion of in signal path and delay, challenging the range as well as time-of-arrival (ToA) estimation. Furthermore, the range error due to beam-squint depends on both range as well as direction information~\cite{delayPhasePrecoding_THz_Dai2022Mar,elbir_MC_beamSquint_DOA_Elbir2023Sep,elbir_thz_SPIM_ISAC_Elbir2023Mar}. Thus, it causes lower (higher) detection and (false alarm) probabilities.     }

	
	\begin{figure*}[t]
		\centering
		{\includegraphics[draft=false,width=.75\textwidth]{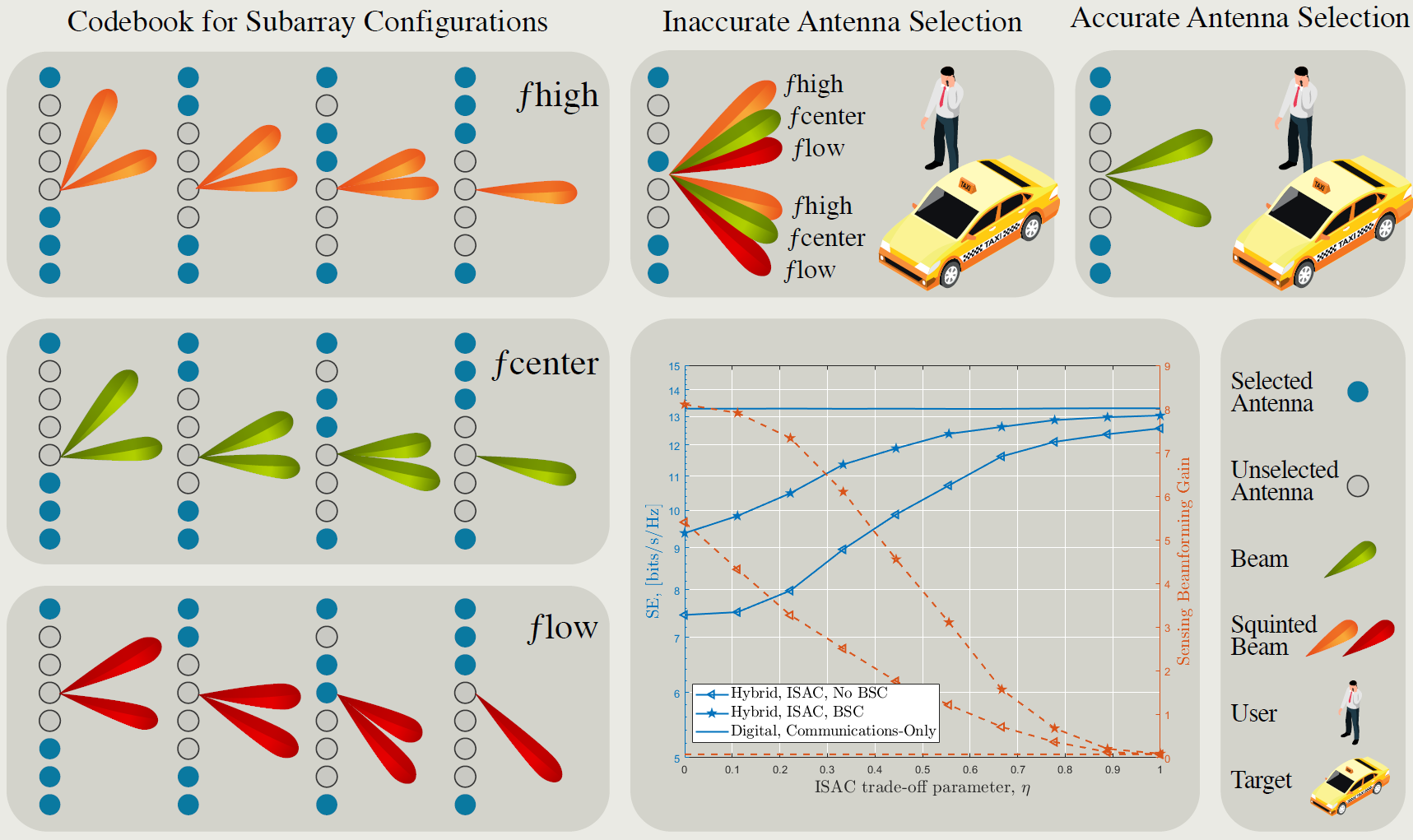} } 
		\caption{ISAC antenna selection involving (left) codebook for subarray configurations at different subcarriers, (top right) illustration of inaccurate/accurate antenna selection, and (bottom right) the communications (SE) and sensing (beamforming gain) performance for selecting $8$ out of $N=128$ when $K=8$, $M=32$ for $300$ GHz carrier frequency and $30$ GHz bandwidth.
		}
		\label{fig_antennaSelection}
	\end{figure*}
	
	\subsection{{Antenna Selection}}
	The ISAC systems necessitate large antenna arrays for high-precision sensing and enhanced capacity with high beamforming gain to compensate for path loss and beam-squint while fewer RF chains are employed for hardware efficiency.  To strike a proper balance, antenna selection is an attractive solution that can select only the high quality subset of antenna elements in the array~\cite{im_ISAC_sparseArray_Wang2018Aug}. However, constructing a non-uniform array raises sidelobe levels and corrupts the array beampattern, which could lead to performance degradation in both S\&C. Thus, antenna selection in ISAC systems is even more challenging than in communications-only and sensing-only systems because it aims to satisfy the performance metrics for both S\&C, and it should take into account the prior information as well as the constraints related to ISAC functionalities.

	An illustration of a codebook of subarray configurations is presented for low/high-end and center frequencies for wideband ISAC scenario is provided in Fig.~\ref{fig_antennaSelection}. The SE performance for ISAC antenna selection is also presented in Fig.~\ref{fig_antennaSelection} with respect to the ISAC trade-off parameter $\eta$ in  beam-squint scenario, wherein $\eta =0$ and $\eta=1$ correspond to radar-only and communications-only implementation, respectively. Once the best subarray configurations are determined by maximizing the SE, the beamforming performance of the selected subarrays are benchmarked with digital array of communications-only scenario. We see that the beamformer with phase correction-based BSC outperforms the randomly selected subarrays as well as the best subarray via brute-force without BSC. That is, it is essential to combine the subarray optimization and beamforming problems by incorporating the BSC techniques, e.g., TTDs or phase correction, for a given user/target configuration. Otherwise, the optimization metric, e.g., SE, signal-to-interference-plus-noise ratio (SINR) or beamforming gain, is maximized at a false directions due to beam-squint, thereby causing the selection of inaccurate subarrays~\cite{im_ISAC_sparseArray_Wang2018Aug,radarCommSurveyfanLiuJSTSP_Zhang2021Sep}. This may pose significant loses in both target detection/estimation for sensing as well as in communications capacity.


	\subsection{{Index Modulation}}
\begin{figure}[t]
		\centering
		{\includegraphics[draft=false,width=.8\columnwidth]{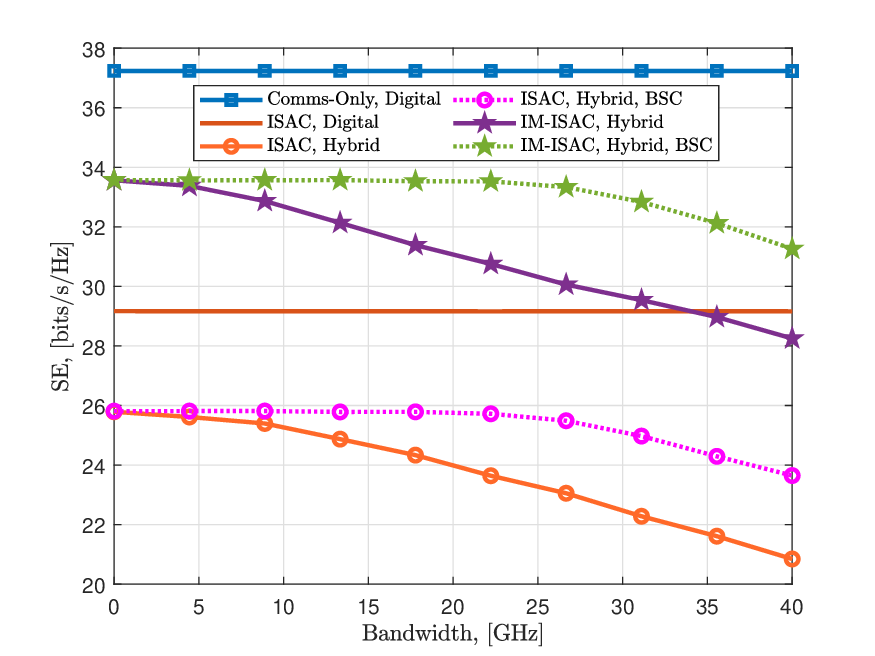} } 
		\caption{SE performance of  IM-empowered ISAC with respect to system bandwidth when $3$ out of $8$ paths are modulated for $\eta = 0.5$, $N=128$, $K=8$, $M=32$, $f_c = 300$ GHz and $\mathrm{SNR} = 0$ dB.
		}
		\label{fig_indexModulation}
	\end{figure}

	IM has been introduced to achieve improved SE than conventional modulation schemes. In IM, the transmitter encodes additional information in the indices of the transmission media such as subcarriers, antennas, spatial paths, etc~\cite{elbir_Beamforming_SPM_Elbir2023Jun}. Thanks to transmitting additional information bits via the indices of activated RF components (e.g., antennas, subcarriers or beamformers), IM-aided systems have shown significant performance improvement in the communications-only systems. In	particular, the IM-aided systems exhibit higher SE than that	the use of fully digital beamformers by transmitting additional encoded information bits into these RF resources. In IM-aided ISAC systems, the BS simultaneously performs IM for communications while generating beams toward the target for sensing. As a result, this performance improvement can be utilized to compensate for the SE loss due to beam-squint. Fig.~\ref{fig_indexModulation} presents the SE performance of conventional and spatial path IM-empowered ISAC beamforming, as illustrated in Fig.~\ref{fig_diag}, with respect to the system bandwidth, wherein the indices spatial paths between the BS and the communications user are considered. Specifically, $3$ out of $8$ paths are used for IM to transmit additional information bits for the ISAC trade-off parameter $\eta = 0.5$. We can see from Fig.~\ref{fig_indexModulation} that the SE performance of ISAC hybrid beamformers degrades as the bandwidth increases because of beam-squint whereas digital beamformers are not affected since they do not include analog components. While BSC based on phase correction provides a relative improvement, IM-empowered ISAC provides much higher SE than that of conventional ISAC thanks to utilizing the spatial path diversity via IM. This observation shows that the SE loss due to beam-squint phenomenon can be alleviated in ISAC systems  by incorporating IM without using complex hardware components.


	
	\section{{Design Challenges}}
	Enabling the aforementioned ISAC applications faces several design challenges including 
	direction and range dependent beam-squint in the near-field, TTD design, array imperfections, waveform design, and beam training. 
	

	\subsection{{Near-field Beam-Squint}}
	The observation of beam-squint differs when the receive array is in the near-field, which is the area where the receive signal wavefront is spherical rather than plane-wave as in far-field. In such a scenario, near-field beam-squint causes the squint of the generated beam toward distinct locations rather than directions.  As a result, handling beam-squint in the near-field is even more challenging than that in the far-field since the impact of beam-squint and the model mismatch due to the spherical wavefront are intertwined. In particular, treating the near-field array response with a far-field plane-wave model constitutes another cause of error, which becomes more problematic as the target/user gets closer to the array~\cite{elbir_Beamforming_SPM_Elbir2023Jun}. Thus, near-field beam-squint brings up new research challenges in ISAC applications, such as target/user DoA estimation,  beamforming, waveform design, and resource allocation. One direct solution, analogous to the far-field methods, is to employ OMP or MUSIC algorithms with the dictionary of near-field array responses for channel/DoA estimation and beamforming~\cite{nearField_ISAC_Wang2023May}. {Unlike far-field, in the presence of near-field beam-squint, the joint design for S\&C requires an accurate computation of the deviation in the array impulse response. Then, the TTD devices can be used to generate time-delays which are controlled with the dependence of both direction and range~\cite{nf_ISAC_mag_Cong2023Oct}.  }


	%
	%
	
	
	\subsection{{TTD Design}}
	A TTD element is expected to introduce constant and tunable time delay response over a wide frequency range.Current TTD circuits based on artificial transmission lines can support a maximum delay range of $500$ ps with $5$ ps delay resolution over $20$ GHz bandwidth~\cite{ttd_Rotman2016Jan,delayPhasePrecoding_THz_Dai2022Mar}. Thus, the applications targeted at wide bandwidths pose a practical challenge for TTD design as the chip area increases exponentially in order to achieve long-time delays. {See~\cite{ttd_mmWave_Lin2022} for a summary of various TTD implementations.} The design of TTD network configuration constitutes another challenging trade-off that should be investigated. For instance, parallel TTD configurations between the RF chains and the phase shifters, as illustrated in Fig.~\ref{fig_diag}, require each TTD to support large delay times with independent control of the TTD units. In contrast, the serial connection of the TTD units facilitates the accumulation of time delays introduced by each serial TTD unit.

	\subsection{{Array Imperfections}}
	In large arrays, the antennas are densely packed, they are composed of hundreds of elements to cope with the severe path loss in high frequency and achieve beamforming gain~\cite{milliDegree_doa_THz_Chen2021Aug}. The dense integration of the antennas is accompanied by mutual coupling (MC), thereby leading to the degradation of array gain~\cite{elbir_MC_beamSquint_DOA_Elbir2023Sep}. Another error source in array design is due to the imperfections of the antenna and cable installments, which cause gain/phase mismatches GPM~\cite{elbir_Beamforming_SPM_Elbir2023Jun}. In practice, the omni-directionality of the array pattern is invalid, hence the impact of both MC and GPM becomes direction-dependent. As a result, beam-squint eventually impacts MC and GPM due to directional beampattern. In comparison, the impact of beam-squint on parameter estimation, e.g., DoA estimation, is more severe than that of MC or GPM. {In Fig.~\ref{fig_DOA_SNR}, DOA estimation RMSE is presented with respect to SNR in the presence of beam-squint and MC~\cite{elbir_MC_beamSquint_DOA_Elbir2023Sep}. We can see that the MC has a marginal impact ($ \sim 0.25^\circ$) on DoA estimation RMSE  whereas the former significantly damages the array gain and causing  $ \sim 4^\circ$ DoA error.} In conclusion, ISAC implementations should consider the intertwined impact of beam-squint on direction-dependent array imperfections. Furthermore,  auto-calibration approaches should be further investigated for the over-the-air calibration of array imperfections for S\&C functionalities.

	{
		\subsection{Carrier Aggregation}
		Carrier aggregation (CA) is an efficient technique to improve the SE in communications by aggregating the fragmented spectrum resources. Also in radar applications, CA is adopted in stepped carrier orthogonal frequency division multiplexing (OFDM) radars and frequency-agile radars to achieve higher range resolution. A CA-aided ISAC setup is devised in~\cite{carrierAggregation_ISAC_Wei2023Oct}, wherein the high and low frequency bands are aggregated to improve the sensing performance. However, beam-squint causes the variation of the array gain across subcarriers, thereby making CA a challenging issue in S\&C, which requires the re-optimization of the CA parameters, e.g., aggregated center frequencies and bandwidths, to account for beam-squint.
		

	}

	\subsection{{Waveform Design and Resource Allocation}}
	The joint processing of S\&C signals in ISAC requires taking into account the trade-off between the S\&C. In particular, the ISAC receiver is responsible for demodulating the communications symbols and recovering the radar echo signals. Furthermore, this resource allocation problem should be solved in a unified manner to improve the integration gain of the ISAC system~\cite{radarCommSurveyfanLiuJSTSP_Zhang2021Sep}. A common technique to allocate power for S\&C tasks is to employ a tuning parameter $\eta$. The selection procedure of $\eta$ involves the ratio of power budgets or the signal duration intervals of S\&C tasks~\cite{elbir_thz_SPIM_ISAC_Elbir2023Mar}. The resource allocation strategy should also take into account the impact of beam-squint, especially for wideband ISAC. The allocation of the S\&C resources over frequency yields a challenge as beam-squint is more dominant over the high/low-end subcarrier frequencies. For instance,  the transmit waveform parameters, e.g., beamforming weights, can be optimized simultaneously with subcarrier allocation to transmit multiple beams toward radar targets and communications users. 
	Consider the waveform design problem in an automotive radar scenario, wherein MIMO arrays are employed for ISAC. In this case, the desired problem requires mutually orthogonal waveforms via optimizing the low integrated sidelobe levels, thereby leading to enhanced target/user detection, high angular/range resolution as well as improved SNR~\cite{elbir_MC_beamSquint_DOA_Elbir2023Sep}. In conclusion, this discussion invites future research studies on waveform design and resource allocation for ISAC in the presence of beam-squint.

	
	\subsection{{Beam Training}}
	In mmWave systems, beam training is performed to align the beamforming directions and realize maximum gain between the BS and the users. A common approach is to assume frequency-flat phase shifters in all antennas to steer or combine the signals toward the direction of interest via beam sweeping~\cite{radarCommSurveyfanLiuJSTSP_Zhang2021Sep}.
	Due to a limited number of RF chains, the beam training stage may be time-consuming for large arrays, which poses a serious constraint for the next generation communications systems. Furthermore, beam-squint should be compensated for accurate beam training with different techniques, e.g., TTD-empowered arrays. To address these fundamental challenges, beam-squint can be exploited to our advantage to provide higher DoF in beam training~\cite{ttd_mmWave_Lin2022,beamSquint_ISAC_Gao2023Mar}. In particular, the TTD units supporting large time delay connected to each RF chain can be used to realize frequency-dependent probing beams, thereby increasing the number of training beams over channel use~\cite{delayPhasePrecoding_THz_Dai2022Mar}. Besides, this approach can also be exploited for sensing in ISAC. That is, the targets can be covered by distinct beams that are generated from different subcarriers thanks to employing a large delay range of the TTDs. Then, based on the echo signal reflected from the targets, which are measured across the subcarriers, the target DoA angles can be obtained~\cite{beamSquint_ISAC_Gao2023Mar}.

	\section{{Summary and Future Outlook}}
	ISAC is perceived to be a promising technology for both high-precision sensing and wireless communications with enhanced capacity in the next-generation wireless networks with exciting applications. This article provided a synopsis of recent developments in ISAC in the presence of beam-squint. The state-of-the-art signal processing techniques are discussed with their advantages/drawbacks along with their performance analysis in various ISAC applications. In particular, TTD-based approaches are advantageous compared to phase shifters thanks to introducing frequency-dependent time delays over a large frequency range whereas phase correction techniques enjoy unneeded additional hardware components at the cost of performance loss dependent on the limited RF chains. 
	
	The impact of beam-squint has been discussed for various ISAC applications. In particular, beam-squint can be easily treated via virtually generated frequency-dependent array responses in channel/DoA estimation tasks which are performed in the digital domain. Nevertheless, applications involving analog processing, e.g., hybrid beamforming, necessitates the implementation of beam-squint compensation techniques. In antenna selection, beam-squint causes the selection of false subarray configurations, which can be overcome by jointly tackling the beam-squint in the optimization problem. IM  is an attractive technique to provide higher SE than that of digital arrays, thereby compensating for the SE loss due to beam-squint.

	The impact of beam-squint in near-field significantly differs from far-field, which suggests the investigation of S\&C applications based on the spherical-wave model. By taking into account beam-squint, certain design stages such as waveform design and beam training should be reorganized. In TTD-based transceivers, beam-squint can be exploited such that frequency-dependent beams are generated in different directions, thereby improving the DoF during training. Nevertheless, this requires a long time delay range of TTD units. Thus, the circuit design for TTD units also poses a challenge for future research. In conclusion, each of the aforementioned scenarios opens up new research directions bringing its fair share of algorithmic and practical nuances.


	\bibliographystyle{IEEEtran}
	\bibliography{references_130}

	\vspace{20pt}
	{\footnotesize
		\textbf{Ahmet M. Elbir} (Senior Member, IEEE)  is a Research Fellow at King Abdullah University of Science and Technology and Duzce University. 
		
		\textbf{Kumar Vijay Mishra} (Senior Member, IEEE)  is a National Academies Harry Diamond Distinguished Fellow at the U. S. Army Research Laboratory.
		
		\textbf{Abdulkadir Celik} (Senior Member, IEEE)  is a Research Scientist at King Abdullah University of Science and Technology.
		
		\textbf{Ahmed M. Eltawil} (Senior Member, IEEE)  is a  Professor at King Abdullah University of Science and Technology.
	}

\end{document}